\documentclass{ieeeaccess}
\usepackage{cite}
\usepackage{amsmath,amssymb,amsfonts}
\usepackage{algorithmic}
\usepackage{graphicx}
\usepackage{textcomp}
\usepackage{nameref,hyperref}
\makeatletter
\renewcommand{\@biblabel}[1]{\quad#1.}
\makeatother

\usepackage{tabularx}   
\usepackage{booktabs}   
\usepackage{bm}
\makeatletter
\AtBeginDocument{\DeclareMathVersion{bold}
\SetSymbolFont{operators}{bold}{T1}{times}{b}{n}
\SetSymbolFont{NewLetters}{bold}{T1}{times}{b}{it}
\SetMathAlphabet{\mathrm}{bold}{T1}{times}{b}{n}
\SetMathAlphabet{\mathit}{bold}{T1}{times}{b}{it}
\SetMathAlphabet{\mathbf}{bold}{T1}{times}{b}{n}
\SetMathAlphabet{\mathtt}{bold}{OT1}{pcr}{b}{n}
\SetSymbolFont{symbols}{bold}{OMS}{cmsy}{b}{n}
\renewcommand\boldmath{\@nomath\boldmath\mathversion{bold}}}
\makeatother

\begin{document}
\history{Date of publication xxxx 00, 0000, date of current version xxxx 00, 0000.}
\doi{10.1109/ACCESS.2024.0429000}

\title{Multi-omic Prognosis of Alzheimer's Disease with Asymmetric Cross-Modal Cross-Attention Network}
\author{\uppercase{YANG Ming}\authorrefmark{1},
\uppercase{JIANG Shi Zhong}\authorrefmark{2}, and \uppercase{ZHOU Su Juan}\authorrefmark{3}
}
\address[1]{College of Medical Information Engineering, Guangdong Pharmaceutical University, Guangzhou, Guangdong 510006, China}
\address[2]{College of Medical Information Engineering, Guangdong Pharmaceutical University, Guangzhou, Guangdong 510006, China}
\address[3]{College of Medical Information Engineering, Guangdong Pharmaceutical University, Guangzhou, Guangdong 510006, China}
\tfootnote{Supported by the National College Students' Innovative and Entrepreneurial Training Program of China (Grant No. 51328064012)}

\markboth
{Author \headeretal: Preparation of  studys for IEEE TRANSACTIONS and JOURNALS}
{Author \headeretal: Preparation of  studys for IEEE TRANSACTIONS and JOURNALS}

\corresp{Corresponding author: \uppercase {ZHOU Su Juan} (e-mail: 616748266@qq.com).}

\begin{abstract}Alzheimer's Disease (AD) is an irreversible neurodegenerative disease characterized by progressive cognitive decline as its main symptom. In the research field of deep learning-assisted diagnosis of AD, traditional convolutional neural networks and simple feature concatenation methods fail to effectively utilize the complementary information between multimodal data, and the simple feature concatenation approach is prone to cause the loss of key information during the process of modal fusion. In recent years, the development of deep learning technology has brought new possibilities for solving the problem of how to effectively fuse multimodal features. This paper proposes a novel deep learning algorithm framework to assist medical professionals in AD diagnosis. By fusing medical multi-view information such as brain fluorodeoxyglucose positron emission tomography (PET), magnetic resonance imaging (MRI), genetic data, and clinical data, it can accurately detect the presence of AD, Mild Cognitive Impairment (MCI), and Cognitively Normal (CN). The innovation of the algorithm lies in the use of an asymmetric cross-modal cross-attention mechanism, which can effectively capture the key information features of the interactions between different data modal features. This paper compares the asymmetric cross-modal cross-attention mechanism with the traditional algorithm frameworks of unimodal and multimodal deep learning models for AD diagnosis, and evaluates the importance of the asymmetric cross-modal cross-attention mechanism. The algorithm model achieves an accuracy of 94.88\% on the test set.
\end{abstract}


\titlepgskip=-21pt

\maketitle

\section{Introduction}
\label{sec:introduction}
\PARstart Globally, over 40 million individuals live with Alzheimer's disease (AD) ~\cite{bib1}, a neurodegenerative disorder characterized by progressive cognitive decline. Current therapeutic strategies for AD predominantly focus on symptomatic management to decelerate disease progression. Mild cognitive impairment (MCI) is widely recognized as the transitional stage between age-related cognitive changes and early dementia, during which timely intervention can preserve optimal cognitive function ~\cite{bib2} and longitudinal monitoring may delay disease progression.

The Alzheimer's Disease Neuroimaging Initiative (ADNI) was established to address critical gaps in AD research. Early studies relied on single-biomarker inputs for deep learning models, whereas recent multimodal deep learning advancements have demonstrated that integrating heterogeneous data modalities consistently outperforms single-modal approaches in diagnostic accuracy ~\cite{bib3}. Clinically, AD's complex etiology often leads to diagnostic ambiguity due to overlapping dementia symptoms with other neurodegenerative disorders. Thus, accurate AD diagnosis necessitates comprehensive integration of multimodal data, including medical imaging (PET/MRI), neurological assessments, and demographic profiles ~\cite{bib4}. Notably, genetic data has emerged as a pivotal diagnostic biomarker, increasingly incorporated into deep learning-based AD frameworks ~\cite{bib5}.

In computer-aided AD diagnosis, early studies focused on single-modal classification  ~\cite{bib6,bib7,bib8} . For example, Zhang et al. ~\cite{bib9} adapted DenseNet's connection architecture to weighted summation for three-class (AD/CN/MCI) classification using MRI, while Basheera et al. ~\cite{bib10} employed skull-stripped MRI inputs in convolutional neural networks. Despite methodological advancements, these single-modal approaches inherently suffer from accuracy limitations due to incomplete pathological feature capture.

Historically, multimodal AD diagnosis relied on simple feature concatenation for cross-modal fusion. Swamy et al. ~\cite{bib11}, for instance, processed MRI using Attention and Inception-based subnetworks before concatenating extracted features with clinical data for classification. Such naive fusion strategies, however, overlook deep-level feature interactions and modality-specific importance, risking loss of critical cross-modal associations and restricting model performance. The challenge of effective multimodal feature fusion to leverage inter-modality complementarity has thus become a central research focus in AD diagnostic modeling.

To address this, attention mechanisms have been introduced to enhance multimodal integration. Zhang et al. ~\cite{bib12} concatenated MRI and PET features prior to cross-attention fusion with cerebrospinal fluid (CSF) biomarkers, while Chen et al. ~\cite{bib13} integrated MRI/PET features with encoded clinical data using channel-spatial attention modules. These approaches, however, are constrained by traditional pairwise cross-attention, which enables only local modal interactions and struggles to capture global associations between imaging (MRI/PET) and non-imaging (clinical/genetic) modalities—highlighting the need for advanced cross-modal fusion frameworks.

Recent research has explored frequency-domain transformers as an alternative to multi-head self-attention to mitigate model parameter complexity. B Patro et al.  ~\cite{bib14}proposed SpectFormer by stacking GFNet with Transformer architecture, demonstrating that integrating frequency-domain transformers with traditional multi-head attention yields superior performance. Inspired by SpectFormer, this study introduces an innovative Asymmetric Cross-Modal Crossover Attention-based Multiclass Diagnostic Framework (ACMCA) for AD. Designed to resolve the challenge of effective multimodal feature fusion, ACMCA integrates frequency-domain transformers with multi-head attention to excavate deep inter-modal feature relationships and enhance joint multimodal representation. The framework employs a parallel Fnet-Transformer architecture, whose efficacy is validated through comprehensive experiments.

This work advances fine-grained AD classification by introducing a trinary diagnostic paradigm (AD/MCI/CN), diverging from early binary (AD/CN) approaches. MCI subjects exhibit pronounced heterogeneity and higher likelihood of AD pathological markers than CN controls, rendering trinary classification more challenging and demanding advanced fine-grained feature learning. Notably, while conventional multimodal studies typically use PET/MRI and clinical data, this research innovatively incorporates genetic data, demonstrating that four-modal integration outperforms three-modal frameworks. Experimental results confirm enhanced diagnostic accuracy by leveraging complementary information across heterogeneous datasets.

\section{Data sources and preprocessing}
\subsection*{A.Dataset description}
The data used in this study were obtained from the Alzheimer's Disease Neuroimaging Initiative (ADNI) database, which provides imaging, clinical, and genetic data for over 2,220 participants across four fundamental research programs (ADNI1, ADNI2, ADNI GO, and ADNI3). Multimodal data from ADNI1, ADNI2, and ADNI GO participants were processed following methodologies reported in previous studies. To effectively integrate diverse modal data, this research focused on participants with available imaging, genetic, and clinical records. For model training, only participants with complete records across all four modalities (referred to as the overlapping dataset) were included, as detailed in Table 1.

\begin{table}[!ht]
\centering
\caption{\textbf{ADNI Dataset Statistics}}
\label{table:adni_statistics}
\begin{tabularx}{\linewidth}{
    >{\centering\arraybackslash}p{2cm}  
    *{4}{>{\centering\arraybackslash}X}  
  }
\toprule  
Data Type & Totals & CN & MCI & AD \\
\midrule  
Clinical Data & 2384 & 942 & 796 & 646 \\
MRI Imaging & 551 & 278 & 123 & 150 \\
PET Imaging & 482 & 225 & 98 & 132 \\
Genetic Data & 805 & 241 & 318 & 246 \\
Data Overlap & 239 & 165 & 39 & 35 \\
\bottomrule  
\end{tabularx}

\begin{flushleft}
\textit{Note: The table shows the number of participants for each type of data and their diagnoses. The following table shows the number of participants and further classifies them into their diagnostic categories. The data overlap refers to testers who had records in all four modalities.}
\end{flushleft}
\end{table}

\subsection*{B. Label and Clinical Data Preprocessing}
During the data analysis and preprocessing, we observed dynamic diagnostic shifts in participants initially categorized as CN or MCI, with some MCI cases progressing to AD longitudinally. Notably, while certain participants underwent up to 16 MRI scans, clinical and genetic data collection frequencies were significantly lower. To address this, we utilized each participant’s most recent clinical profiles and updated diagnostic labels.

For clinical data, we included neurological assessments and demographic records from 2,384 participants. Neurological metrics reflect AD-related impacts on distinct central nervous system regions ~\cite{bib15}, while demographic factors—age and gender—have been shown to influence brain physiology ~\cite{bib16}. Integrated clinical datasets comprised seven variables: gender, age, Montreal Cognitive Assessment (MoCA), Mini-Mental State Examination (MMSE), Clinical Dementia Rating (CDR), Functional Activities Questionnaire (FAQ), and Geriatric Depression Scale (GDS).

\subsection*{C. Imaging Data Preprocessing}
Previous studies ~\cite{bib17} have demonstrated that MRI can clearly reflect disease progression in AD participants even at early stages. Amyloid plaques, one of the primary neuropathological markers of Alzheimer's disease ~\cite{bib18}, form in close relation to gene expression ~\cite{bib19}. Amyloid PET imaging using 18F-AV-45 as a tracer enables noninvasive in vivo detection of these plaques. To minimize model parameters and computational resource consumption, this study selected mid-sagittal slices from participant MRI and PET scans, as illustrated in Figures 1 and 2.

\subsection*{D. Gene Data Preprocessing}
The genetic data in this study were derived from ADNI, comprising whole-genome sequencing (WGS) data of 805 participants. Sequencing was completed by Illumina's non-CLIA laboratory during 2012–2013, with raw variant call format (VCF) files generated by ADNI using the Burrows-Wheeler Aligner (BWA) and Genome Analysis Toolkit (GATK). Genomic data preprocessing followed established protocols detailed in Reference ~\cite{bib20}, involving:
\textbf{SNP Filtration:}  Raw SNPs were filtered using Hardy-Weinberg equilibrium (HWE) testing ($p < 0.05$), genotype quality (GQ $\geq$ 20), minor allele frequency (MAF $\geq$ 0.01), and missing rate ($\leq$ 0.05).
Gene-Specific SNP Enrichment: Six hundred eighty AD-related genes were obtained from the AlzGene database, of which 640 were retained after annotation matching via the UCSC Genome Browser and NCBI RefSeq. This yielded 547,863 SNPs within these gene regions.
Dimensionality Reduction: Supervised feature selection using a random forest classifier reduced the feature space to 15,000 SNPs. Final alleles were categorized into three states: no allele, single allele, and two alleles based on allelic presence.

\subsection*{E. Determination of Training Datasets}
The overlapping dataset for AD diagnostic type prediction was derived from raw data sources—551 participants with MRI, 482 with PET, 805 with single-nucleotide polymorphism (SNP) profiles, and 384 with complete clinical records. Through intersectional filtering to retain only participants with multimodal data availability, a final cohort of 239 participants was selected for training the ACMCA model. Demographic characteristics of the cohort are summarized in Table 2.

\begin{table}[!ht]
\centering
\caption{\textbf{Demographic Information}}
\label{table:demographic_info}
\begin{tabularx}{\linewidth}{
    >{\centering\arraybackslash}p{2cm}X  
    *{3}{>{\centering\arraybackslash}X}  
  }
\toprule  
Group & Totals & Female (\%) & Age (yr) \\
\midrule  
CN & 165 & 53.9 & 77.8 \\
MCI & 39 & 34.2 & 76.6 \\
AD & 35 & 31.4 & 78.1 \\
\bottomrule  
\end{tabularx}

\begin{flushleft}
\textit{Note: This table shows the number of participants in each category, as well as the proportion of females and the mean age in each group.}
\end{flushleft}

\end{table}

\Figure[t!](topskip=0pt, botskip=0pt, midskip=0pt){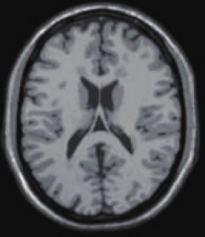}
{ \textbf{MRI Image Slice ID: 012\_S\_6760 . }\label{fig1}}

\Figure[t!](topskip=0pt, botskip=0pt, midskip=0pt){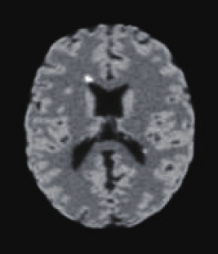}
{ \textbf{PET Image Slice ID: 012\_S\_6760 . }\label{fig2}}

\section{Asymmetric Cross-Modal Cross-Attention (ACMCA) network}
This  study constructs the ACMCA model to fully explore the cross-modal data association information among MRI images, PET images, gene data, and clinical data for the diagnostic task of three types of AD.
The ACMCA model consists of four major modules: a shallow feature extraction module, a multimodal feature fusion module, a deep feature extraction module, and an output classification module. In the shallow feature extraction module, due to the characteristics of different modal data, this  study adopts a differentiated feature extraction method, which is mainly divided into a medical image feature extraction module and a structural data feature extraction module. After completing the shallow feature extraction, the extracted features are input into the multimodal feature fusion module. In this module, an asymmetric cross-modal cross-attention mechanism is used to achieve the fusion of four types of data features and generate fused multimodal features. These fused features are not directly used for classification but are input into the deep feature extraction module composed of a self-attention mechanism and a frequency domain transformer. In this deep feature extraction module, the model can aggregate multimodal feature information, thereby deeply learning the deep connections between different modal features, and then input the feature information into the classification module. The specific model architecture is shown in Figure 3.
\Figure[t!](topskip=0pt, botskip=0pt, midskip=0pt)[width=1\linewidth]{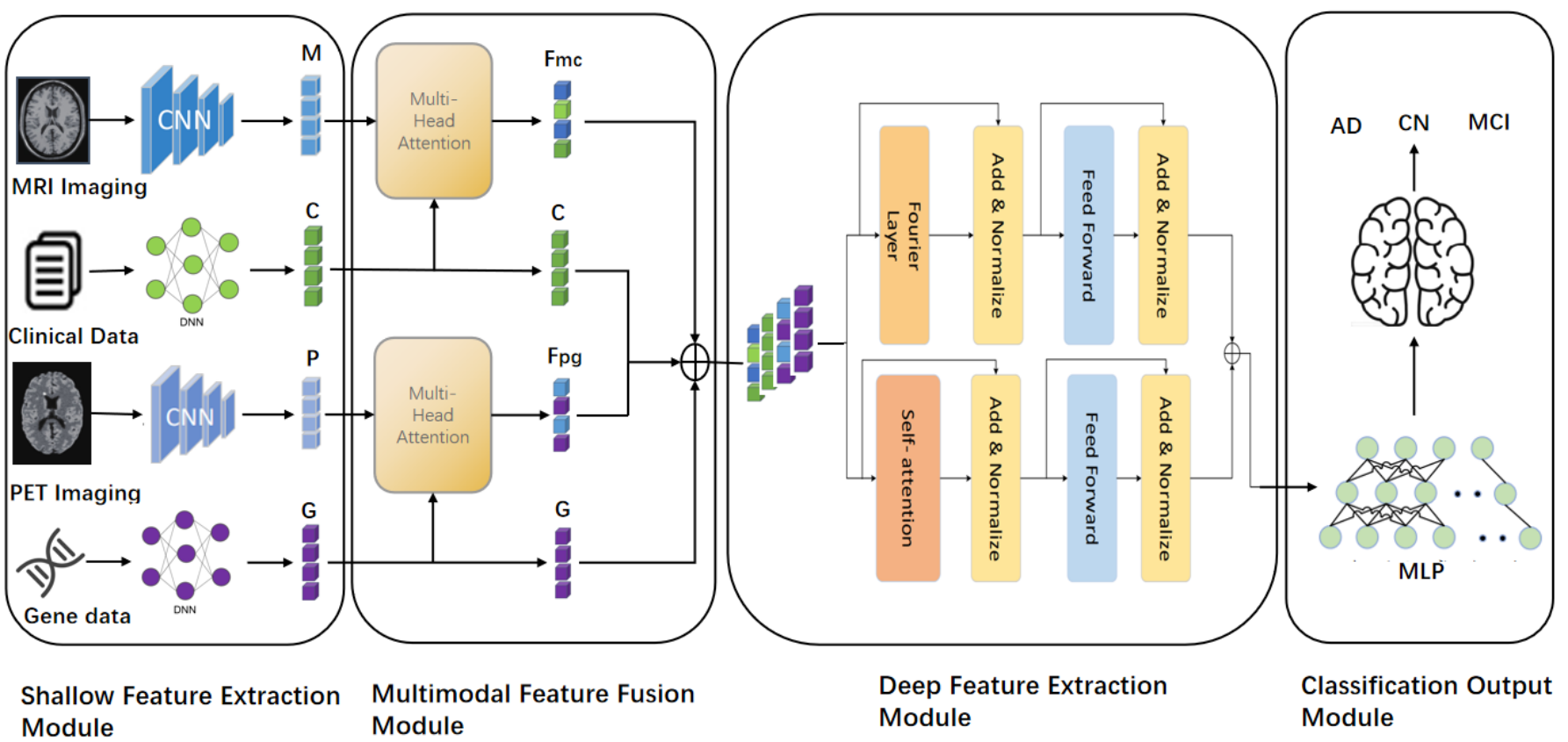}
{ \textbf{Famework of ACMCA. }\label{fig3}}

\subsection*{A.Shallow Feature Extraction Module}
This  study uses the pre-trained model ResNet50 for feature extraction of image modality to obtain richer image feature information. Since MRI and PET images are medical imaging data, ResNet50 can effectively extract high-level image semantic features from them. For genetic data, three types of data (no allele, any one allele, and both alleles present) are converted into the input format of a Deep Neural Network (DNN), represented as encoded vectors, and dimensionality reduction operations are performed synchronously. For clinical data (including 7 detection indicators of testers), DNN is used to extract features from the encoded clinical data, and dimensionality enhancement operations are performed synchronously. This ensures that the features of clinical data, genetic data, MRI images, and PET images are consistent in dimension. Unifying the dimensions of these features can make them have similar representations in the feature space, providing a foundation for the subsequent feature fusion module.
\subsection*{B.Multimodal Feature Fusion Module}
\Figure[t!](topskip=0pt, botskip=0pt, midskip=0pt )[width=6 in]{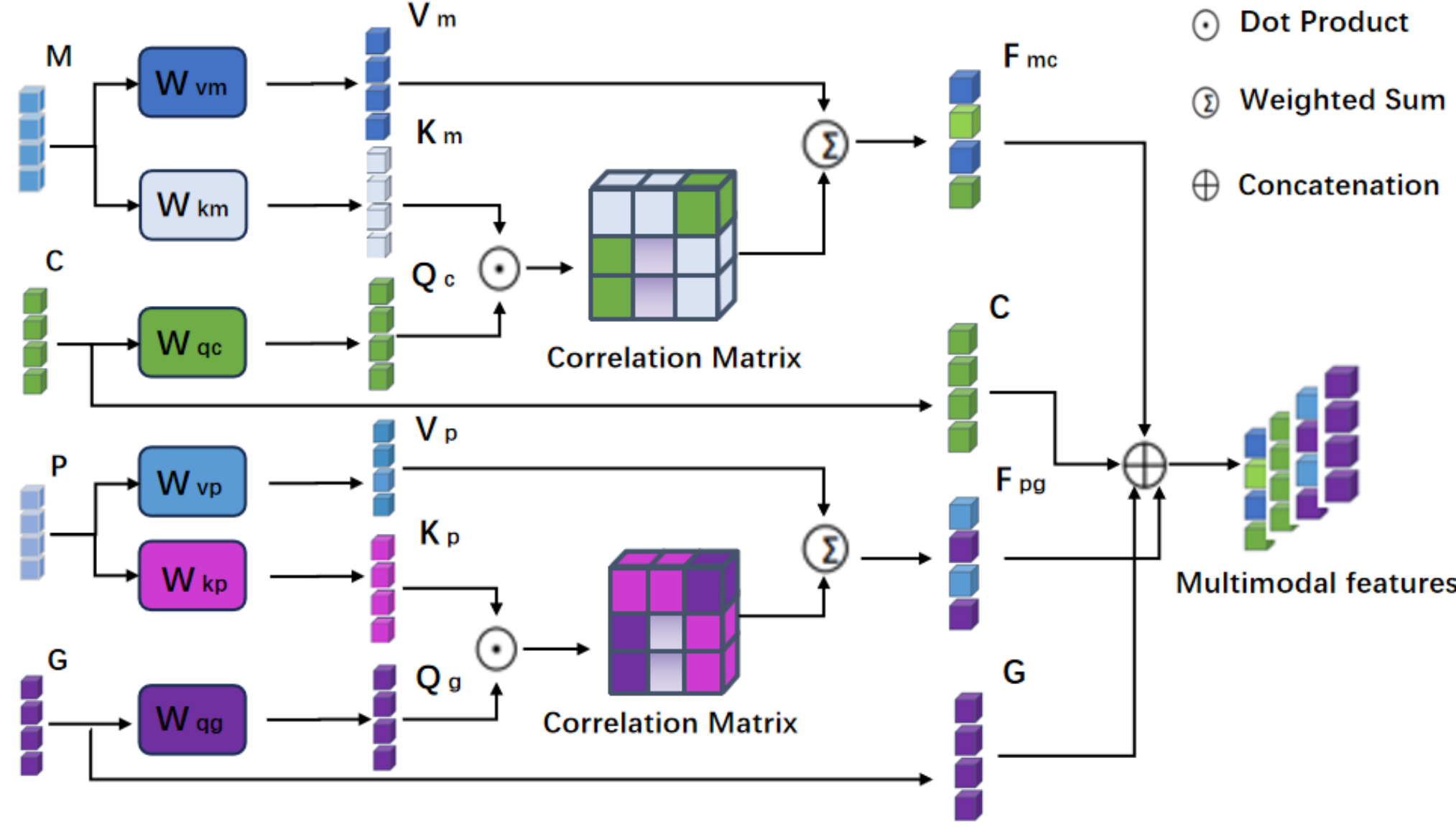}
{ \textbf{ Famework of ACM. }\label{fig4}}
Since clinical data and genetic data, as structured data, have significant differences in data structure and feature distribution from medical imaging data, more attention should be paid to the significant differential relationships between different modalities when fusing cross-modal data. This  study proposes an asymmetric cross-modal cross-attention mechanism to capture the cross-modal information interaction details between clinical data, genetic data, MRI images, and PET images. The specific module design is shown in Figure 4. This mechanism maps the features of different modalities into Query, Key, and Value vectors through position encoding. The correlation between positions is determined by calculating the similarity between the query vector and the key vector. Finally, the value vectors are weighted and summed to obtain a representation vector containing attention information from different modalities, achieving effective fusion of cross-modal information.

In the multimodal fusion process of the ACMCA model, the fusion of clinical data, genetic data, MRI images, and PET images is called the Image-numerical fusion branch. The operation process in this fusion branch is based on specific principles and formulas. First, the clinically data features and genetic data features after position encoding are processed to become query vectors, which follows Formula (1) and Formula (2). The MRI features and PET features are respectively mapped into corresponding key vectors and value vectors, corresponding to Formula (3) and Formula (4). Next, for each query vector, the dot product with all key vectors is calculated, divided by the square root of the vector dimension, and then the Softmax function is used to convert the result into a probability vector. Through this operation, the correlation matrix between features at different positions can be obtained. Finally, the correlation matrix is used to calculate with the value vector to obtain the image representation processed by the attention mechanism, which can be represented by Formula (5) and Formula (6).
\begin{align}
Q_c &= W_{qc}C \tag{1} \\
Q_g &= W_{qg}G \tag{2} \\
K_m &= W_{km}M \quad, \quad V_m = W_{vm}M \tag{3} \\
K_p &= W_{kp}P \quad, \quad V_p = W_{vp}P \tag{4} \\
F_{mc} &= \text{CMA}(Q_c, K_m, V_m) = \text{Softmax}\left(\frac{Q_c K_m^T}{\sqrt{d_k}}\right) V_m \tag{5} \\
F_{pg} &= \text{CMA}(Q_g, K_p, V_p) = \text{Softmax}\left(\frac{Q_g K_p^T}{\sqrt{d_k}}\right) V_p \tag{6}
\end{align}   
In the multimodal fusion module of this  study, the features output from the clinical data feature extraction module are denoted as C, which contains the tester's clinical data, neuropsychological test results, demographic data, and other information. Additionally, in this module, the features extracted from genetic data are denoted as G, M represents the features extracted from MRI images, and P denotes the features obtained from PET images. The feature dimensions of C, M, G, and P are all set to 100, meaning they are all 100-dimensional vectors for subsequent feature fusion.During the training process of the ACMCA model, W plays a crucial role as an iteratively optimizable transformation matrix. In the model training stage, W continuously adjusts its matrix values to achieve linear transformations of features from different modalities. Within the framework of the attention mechanism, \(Q_c\) and \(Q_g\) are query vectors obtained by linearly transforming C and G through \(W_{qc}\) and \(W_{qg}\). Their role is to learn relevant feature information from the features of MRI and PET images, assisting the model in acquiring key cross-modal data information.The MRI image features M and PET image features P are respectively converted into key vectors \(K_m\), \(K_p\) and value vectors \(V_m\), \(V_p\) through transformation matrices. The key vectors are used by the query vectors to calculate similarity for determining the degree of association between features at different positions, while the value vectors undergo weighted summation based on the similarity calculation results to finally generate multimodal fusion features. These multimodal fusion features contain association information between different modalities.Among them, \(F_{mc}\) is the feature obtained by fusing MRI image features and clinical data features, and \(F_{pg}\) is the feature obtained by fusing PET image features and genetic data features. After a series of complex and ordered calculations, the output of this module is represented by Equation (7):
\begin{align}
\text{concat}\left(F_{mc}, C, F_{pg}, G\right) \tag{7}
\end{align}
This fusion method based on asymmetric cross-modal cross-attention can guide the model to focus on learning the alignment relationship between medical images and numerical data, thereby efficiently understanding the internal connection between image data blocks and clinical data. The mixed features produced by this fusion method are defined as multimodal features, with their tensor form being (batch size, number of channels, feature dimension), specifically (32, 3, 100).

\subsection*{C. Deep Feature Extraction Module}
The deep feature extraction module is responsible for deeply extracting the fused features output by the multimodal attention module. Its role is to extract the effective correlation information within the multimodal features, thereby enabling better interaction of features at different positions. The module is composed of a self-attention module and a Fourier module. Researchers such as Patro B N have combined the transformer with a frequency domain transformer ~\cite{bib14}, using frequency domain analysis to enhance the feature extraction effect. Their research has shown that this combination can strengthen the model's feature extraction performance and effectively improve the model's accuracy, which will also be verified in the experimental section of this  study.

The Fourier module uses the Fnet structure and applies discrete Fourier transform in the Fourier Layer to transmit the internal information of multimodal features. For a sequence of length N, the formula for the fast Fourier transform is:
\begin{align}
X[k] &= \sum_{n=0}^{N-1} e^{-i\frac{2\pi}{N}nk} \ x[n] \quad k=0,1,\ldots,N-1 \tag{8}
\end{align}
Specifically, for each position k of the feature, the Fnet structure performs a Fourier transform on the original multimodal features along the dimensional direction, converting them into a new representation. This new form incorporates information from other positions of the feature. By transforming the global information of the sequence into the frequency domain, the model can capture the relationships between different positions in the sequence. Similar to the Transformer architecture, after the frequency-domain representation is output, it enters a fully connected feedforward neural network. Additionally, residual connections and layer normalization operations are performed after both the Fourier transform layer and the feedforward neural network layer.
\subsection*{D.Output Classification Module}
The features output by the deep feature extraction module are fed into the classification module, which is designed with a four-layer multi-layer perceptron (MLP). First, a fully connected layer performs linear transformation on the deep multimodal features, followed by the ReLU activation function to introduce non-linear transformation. Finally, three classification results are output: CN, MCI, and AD. During model training, Cross Entropy Loss is used as the loss function, defined as:
\begin{align}
L &= \frac{1}{N}\sum_{i} L_i = -\frac{1}{N}\sum_{i}\sum_{c=1}^{M} y_{ic}\log\left(p_{ic}\right) \tag{9}
\end{align}  
Here, i represents the total number of samples, and M denotes the total number of categories in the classification task. In view of the fact that this  study focuses on the three-class diagnostic task of Alzheimer's disease, the value of M is set to 3 here. Among them, y corresponds to the true labels of the samples, intuitively reflecting the actual categories to which the samples belong; p is the predicted value output by the model, reflecting the model's judgment results on the categories of the samples.By continuously reducing the value of this loss function, the parameters of the model are optimized to make the prediction results of the model closer to the real situation, so as to improve the performance of the model in the AD classification task.

\section*{Experiments and Results}
\subsection*{A.Hyper parameter search and optimizat}
To determine the optimal parameter combination for model training, this  study conducted hyperparameter experiments. During the experiments, certain parameters were fixed while investigating the impacts of different training epochs, batch sizes, and feature dimensions on the final classification accuracy of the model. When the batch size was set to 32 and the feature dimension was 100, the changes in classification accuracy under different training epochs were observed, and the experimental results are presented in Figure 5. It can be clearly seen from Figure 5 that as the number of training epochs increases, the classification accuracy of the model first rises and then declines.

\Figure[t!](topskip=0pt, botskip=0pt, midskip=0pt){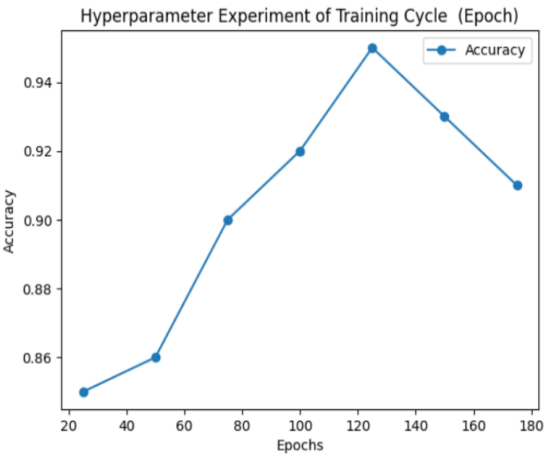}
{ \textbf{  Hyperparameter experiment of training cycle. }\label{fig5}}

In another set of experiments, the training epochs were fixed at 125 and the feature dimension remained at 100 to investigate the impact of different batch sizes on the model's classification accuracy. The relevant results are shown in Figure 6. Through the analysis of Figure 6, the impact of batch size on the model's accuracy can be intuitively understood.
\Figure[t!](topskip=0pt, botskip=0pt, midskip=0pt){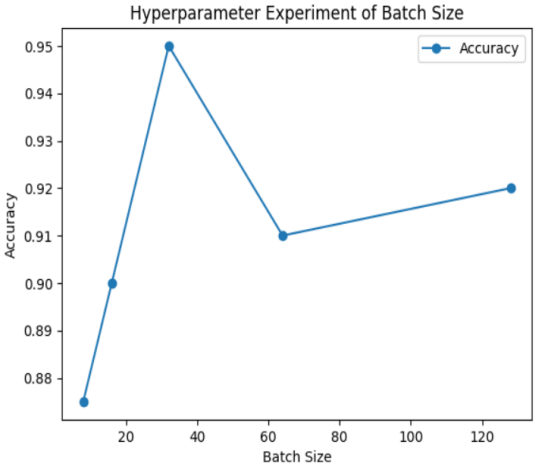}
{ \textbf{  Hyperparameter experiment of batch size. }\label{fig6}}

Meanwhile, experiments on feature dimensions were also conducted. Under the conditions of a training epoch of 125 and a batch size of 32, the impact of different feature dimensions on the model's classification accuracy was studied, and the experimental results are shown in Figure 7. The changing trends of the model's classification accuracy under different feature dimensions can be observed from Figure 7.

\Figure[t!](topskip=0pt, botskip=0pt, midskip=0pt){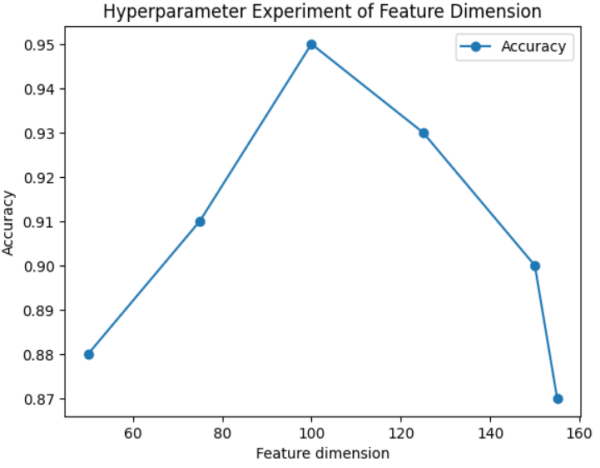}
{ \textbf{  Hyperparameter experiment of feature dimension. }\label{fig7}}

By comprehensively comparing the performance of the model under various hyperparameter settings, the optimal set of hyperparameters was screened out for training the model in this  study. The hyperparameters used for the final trained model are shown in Table 3 below.
\begin{table}[!ht]
\centering
\caption{\textbf{Best Parameter Settings for the Model}}
\begin{tabularx}{\linewidth}{
    >{\centering\arraybackslash}p{5cm}  
    *{1}{>{\centering\arraybackslash}X}  
  }
\toprule  
Parameter Name & Value \\
\midrule  
Imaging Feature Dimension & 100 \\
Non-Imaging Feature Dimension & 100 \\
Image Size & 224×224 \\
Learning Rate & 1e-3 \\
Batch Size & 32 \\
Number of Epochs & 125 \\
\bottomrule  
\end{tabularx}
\label{table:best_parameters}
\end{table}

\subsection*{B.Evaluation Metrics}
To evaluate the performance of the ACMCA model in the multi-class prediction task of Alzheimer's disease (AD), this study conducts experimental comparisons based on MRI images, PET images, genetic data, and clinical data from the ADNI database, and compares it with other baseline models. In terms of performance evaluation, four typical metrics are selected: Accuracy, Recall, Specificity, and F1-score. The calculation methods for each metric are as follows:
\begin{align}
\text{Accuracy} &= \frac{TP + TN}{TP + TN + FP + FN} \tag{10} \\
\text{Recall} &= \frac{TP}{TP + FN} \tag{11} \\
\text{Specificity} &= \frac{TN}{TN + FP} \tag{12} \\
F_1 &= \frac{2 \times TP}{2 \times TP + FP + FN} \tag{13}
\end{align}
In the model evaluation system of this  study, True Positive (TP) represents the number of positive samples correctly identified by the model; True Negative (TN) is the number of negative samples correctly identified by the model; False Positive (FP) refers to negative samples misclassified as positive; False Negative (FN) refers to positive samples misclassified as negative. To systematically evaluate the model performance, the Receiver Operating Characteristic (ROC) curve is used as the evaluation tool. This curve intuitively demonstrates the discriminative performance of the classifier by dynamically depicting the relationship between the True Positive Rate (TPR, i.e., sensitivity, reflecting the model's ability to identify positive samples) and the False Positive Rate (FPR, which equals 1 minus specificity, reflecting the model's identification accuracy for negative samples). Meanwhile, the Area Under the ROC Curve (AUC) is used to quantitatively analyze the classifier's performance. This indicator comprehensively reflects the model's classification performance under different thresholds. The average ROC curve can reflect the model's sensitivity and specificity, and a larger area under the curve indicates more excellent comprehensive performance of the model in identifying both positive and negative samples.

\subsection*{C.Multimodal Effectiveness Experiments}
To investigate whether multimodal data play a role in providing complementary information for AD diagnosis, this  study conducted a series of experiments based on different numbers of modalities to perform three-class classification for AD, MCI, and CN.

In the single-modality experiments, MRI and PET images were tested separately. The ResNet50 model was used to extract features from the image data. Instead of passing through the multimodal attention module, the extracted image features were directly input into the deep feature extraction module. Through this operation, a single-modality feature representation with self-attention weight information was obtained. This single-modality feature was then input into the classification module to complete the AD classification task. When only clinical data or genetic data were used, the data were first encoded, and then a deep neural network (DNN) was used to extract features. The subsequent process was consistent with the experiments using only image modality data. The experimental results are summarized in Table 4. During the single-modality experiments, the experiment relying solely on genetic data showed the most prominent performance. This phenomenon suggests that the genetic data containing multiple SNPs gene expression data of testers contains more diverse information compared to single two-dimensional medical image data.

\begin{table}[!ht]
\centering
\caption{\textbf{Comparison of Modal Performance}}
\begin{tabularx}{\linewidth}{
    >{\centering\arraybackslash}p{2cm}  
    *{4}{>{\centering\arraybackslash}X}  
  }
\toprule
Modality & Accuracy & Recall & Specificity & F1-Score \\
\midrule
\bfseries Genetic Data & \bfseries 0.838 & \bfseries 0.712 & \bfseries 0.894 & \bfseries 0.791 \\
Clinical Data & 0.810 & 0.553 & 0.853 & 0.773 \\
MRI & 0.823 & 0.551 & 0.843 & 0.562 \\
PET & 0.834 & 0.637 & 0.881 & 0.673 \\
Clinical Data + MRI & 0.845 & 0.809 & 0.912 & 0.824 \\
Genetic Data + PET & 0.839 & 0.794 & 0.903 & 0.832 \\
\bfseries Clinical Data + MRI + Genetic Data + PET & \bfseries 0.865 & \bfseries 0.821 & \bfseries 0.923 & \bfseries 0.872 \\
\bottomrule
\end{tabularx}
\label{table:modality_performance}
\end{table}

Compared with using single-modal data, leveraging multimodal data for AD diagnosis demonstrates superior performance. When combining clinical data with MRI image data, the diagnostic accuracy can reach 84\% - 85\%, and the F1 score exceeds 82\%, both of which outperform the results obtained from all single-modal experiments. Similarly, integrating genetic data with PET image data yields a diagnostic accuracy of 83.9\% and an F1 score exceeding 83\%, again surpassing all single-modal counterparts. Notably, the model achieves optimal classification performance when simultaneously utilizing all four modalities: genetic data, clinical data, MRI images, and PET images. This clearly indicates that biomarkers from different modalities can capture disease-related information from diverse perspectives, and their complementary nature significantly enhances the model's classification capabilities.

\subsection*{D. Performance Comparison Experiments}
To verify the excellent performance of the ACMCA model in AD diagnostic classification, this study compares and analyzes it with various baseline models in recent years. The following is a brief introduction to each baseline model:

Model 1: J Venugopalan et al. ~\cite{bib20} integrated extracted imaging and non-imaging features by simple concatenation, and then directly used a random forest classifier for AD multi-classification diagnosis.

Model 2 (MCAD): J Zhang et al. used a cross-attention mechanism to fuse imaging and non-imaging features, performed dimensionality reduction on the fused features, and optimized the model network architecture with cross-entropy loss and modality alignment loss.

Model 3: H Chen et al. successively input imaging features and non-imaging features into a channel attention module and a spatial attention module for modality fusion. The fused features were further extracted by a convolutional module and finally fed into a classification head.

Model 4 (MADDi): M Golovanevsky et al. ~\cite{bib21} enabled three different modal features to first pass through a multi-head self-attention module, then performed cross-attention operations on the processed features in pairs, and finally concatenated the fused features for classification.

Model 5 (MADDi-ACM): This model replaces the multimodal fusion method in MADDi with the asymmetric cross-attention mechanism of this study.

Model 6 (ACMCA-CM): This model replaces the asymmetric cross-attention module in the MAMDF model with a symmetric cross-attention module.

Model 7 (ACMCA-MCAD): Imitating the method of J Zhang et al., this model first concatenates MRI and PET imaging features, and then uses a cross-attention mechanism to fuse non-imaging features.

Model 8 (ACMCA): The cross-modal cross-attention mechanism model in this  study. The experimental results of comparing this model with the above baseline models are shown in Table 5.

\begin{table}[!ht]
\centering
\caption{\textbf{Experimental Results of Performance Comparison}}
\begin{tabularx}{\linewidth}{
    >{\centering\arraybackslash}p{2cm}  
    *{4}{>{\centering\arraybackslash}X}  
  }
\toprule  
Model & Accuracy & Recall & Specificity & F1-Score \\
\midrule  
Model 1 & 0.780 & 0.780 & 0.770 & 0.780 \\
MCAD & 0.735 & 0.639 & 0.820 & 0.619 \\
Model 3 & 0.841 & 0.847 & 0.847 & 0.812 \\
MADDi & 0.859 & 0.761 & 0.924 & 0.769 \\
MADDi-ACM & 0.885 & 0.798 & 0.934 & 0.813 \\
ACMCA-CM & 0.865 & 0.810 & 0.933 & 0.810 \\
ACMCA-MCAD & 0.859 & 0.787 & 0.928 & 0.793 \\
ACMCA & 0.948 & 0.865 & 0.935 & 0.855 \\
\bottomrule  
\end{tabularx}
\label{table:performance_comparison}
\end{table}

\Figure[t!](topskip=0pt, botskip=0pt, midskip=0pt)[width=3 in]{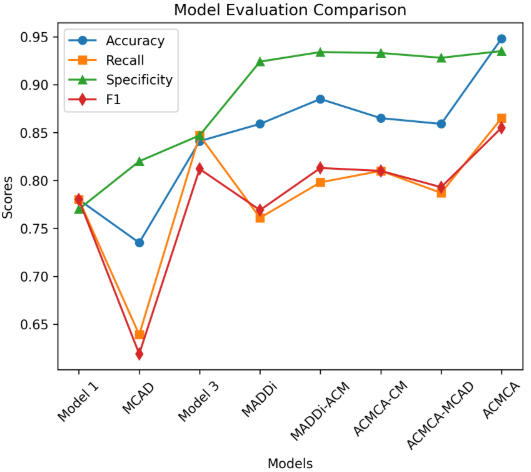}
{ \textbf{  Model comparison experiments. }\label{fig8}}

The statistical results from Table 4 and Figure 8 demonstrate that the attention mechanism can significantly enhance the model's performance in multi-classification tasks. Model 1 constructed by J Venugopalan et al. simply concatenates features from different modalities. This rough feature extraction method makes it difficult for the model to learn the correlation information between multimodal features. In the performance comparison experiment of this  study, the MADDi model, Model 3 developed by H Chen et al., and the proposed ACMCA model all use the attention mechanism to process features of different modalities. Compared with the experimental results of Model 1, these models have achieved improvements in classification accuracy of 16.8\% (ACMCA vs. Model 1), 6.1\% (Model 3 vs. Model 1), and 7.9\% (MADDi vs. Model 1), respectively. The improvement in accuracy fully shows that using the attention mechanism in the modality fusion process can better learn the internal correlation between multimodal feature data.

The experimental results in Figure 8 clearly show that the asymmetric cross-modal cross-attention mechanism adopted by the ACMCA model has significant advantages in fusing imaging features and non-imaging features. When performing multimodal fusion, the MADDi model uses a method of directly performing symmetric cross-attention processing on two modalities and then concatenating features. When the symmetric cross-attention mechanism of the MADDi model is replaced with the asymmetric cross-modal cross-attention mechanism of this study, the model's attention to the connection between imaging features and non-imaging features is significantly improved, and the classification accuracy is also improved by 2.6\%. J Zhang et al. also recognized the strong heterogeneity between imaging features and non-imaging features. In Model 2, they first concatenate the imaging features and then perform symmetric cross-attention processing on the concatenated imaging features and non-imaging features. In this study, after replacing the original asymmetric cross-modal cross-attention mechanism of the ACMCA model with the same processing method as Model 2, it is found that the classification performance of the model has decreased to a certain extent.

\subsection*{E. Ablation Experiments}
To validate the effectiveness of each substructure of the model, in this section of experiments, the ACMCA model proposed in this  study was compared with four ablated models on the same overlapping dataset for the AD three-class classification task. The four ablated models are as follows:

ACMCA-WCM: This model is based on the ACMCA model but removes the cross-attention module, allowing imaging and non-imaging features to directly enter the deep feature extraction module.

ACMCA-WDE: In this variant, the features output by the asymmetric cross-modal cross-attention fusion module are concatenated (removing the deep feature extraction module), and then directly fed into the classification task.

ACMCA-WFnet: This model masks the Fourier module in the deep feature extraction module of the ACMCA model while keeping other modules unchanged.

ACMCA-WT: Derived from the ACMCA model, this variant masks the self-attention module in the deep feature extraction module while maintaining the rest of the architecture.

\begin{table}[!ht]
\centering
\caption{\textbf{Table 6  Results of Ablation Experiments}}
\begin{tabularx}{\linewidth}{
    >{\centering\arraybackslash}p{2cm}  
    *{5}{>{\centering\arraybackslash}X}  
  }
\toprule  
Model & Accuracy & Recall & Specificity & F1 & AUC \\
\midrule  
ACMA-WCM & 0.785 & 0.724 & 0.932 & 0.722 & 0.910 \\
ACMA-WDE & 0.805 & 0.751 & 0.939 & 0.751 & 0.927 \\
ACMA-WFnet & 0.825 & 0.778 & 0.945 & 0.776 & 0.926 \\
ACMA-WT & 0.826 & 0.782 & 0.946 & 0.778 & 0.938 \\
ACMCA & 0.946 & 0.865 & 0.935 & 0.855 & 0.948 \\
\bottomrule  
\end{tabularx}
\label{table:ablation_results}
\end{table}

The experimental results in Table 6 show that the performance of the ACMCA model declines significantly after removing the asymmetric cross-modal cross-attention module. This phenomenon fully demonstrates that the asymmetric cross-modal cross-attention mechanism can effectively explore the internal relationships between different modal data, thereby improving the model's diagnostic classification performance. Compared with the original ACMCA model, removing the deep feature extraction module also leads to a decline in model performance. In addition, when masking the Fnet module and Transformer module in the deep feature extraction (DE) module respectively, the final experimental results show that the performance of the two models is relatively similar, confirming that the frequency domain converter can play a role similar to the self-attention mechanism.

\Figure[t!](topskip=0pt, botskip=0pt, midskip=0pt)[width=3 in]{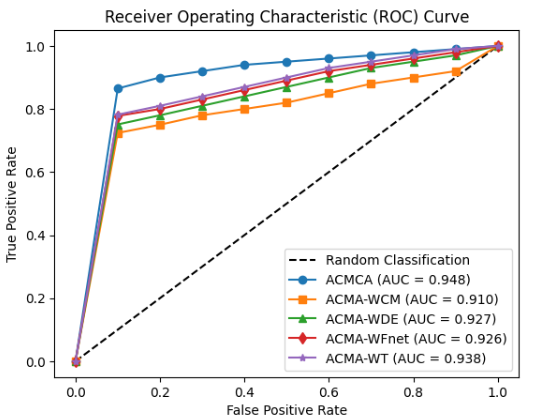}
{ \textbf{  ROC curves for ACMCA ablation experiment. }\label{fig9}}

Figure 9 clearly shows the ROC curves of each model in the ablation experiment. The area under the ROC curve can comprehensively reflect the classification ability of the model. Compared with the original model, removing the deep feature extraction module reduces the AUC value by 0.021 in the AD multi-classification task; removing only the self-attention module decreases the AUC value by 0.01; and removing only the Fourier module reduces the AUC value by 0.022. These data fully indicate that the deep feature extraction module cannot achieve ideal results simply through network stacking, and the rationality of its design is extremely critical.

\section*{5 Conclusion}
Alzheimer's disease (AD) primarily affects brain functions and is commonly seen in the elderly. As the disease progresses, testers may gradually lose their ability to take care of themselves, making treatment particularly complex and challenging. Since there is currently no specific drug for AD, early diagnosis and drug intervention are of great importance, as they can effectively delay the progression of the disease. Aiming at the classification problem of AD, this  study proposes a deep learning network framework based on an asymmetric cross-modal cross-attention mechanism. The framework uses the asymmetric cross-attention mechanism to enhance the associative learning ability between imaging and non-imaging features of different modalities. When processing multimodal medical data, the model excellently completes the tasks of multimodal data fusion and classification output.

The main work and contributions of this  study can be divided into the following three aspects:

    (1) Innovatively combining four types of medical multi-perspective data with different modalities, including PET images, MRI images, clinical data, and genetic data, to train a prediction model for AD.
    
    (2) Proposing a deep learning network model with an asymmetric cross-modal cross-attention mechanism in the feature fusion module to strengthen the associative learning ability between features of different modalities.
    
    (3) Combining the Transformer architecture with a frequency domain converter in the deep feature extraction module, and using frequency domain analysis to improve the effect of feature extraction.

\section*{Future work}
The main purpose of this  study is to enhance the classification and prediction capabilities of deep learning networks for AD. Aiming at this goal, this  study proposes a network architecture based on an asymmetric cross-modal cross-attention mechanism. The classification performance is improved by utilizing multi-modal data, improving the fusion ability between cross-modal data, and enhancing deep feature extraction capabilities. The effectiveness of the model is fully verified through extensive correlation experiments. However, there are still limitations and research directions worthy of in-depth exploration:

(1) The network model proposed in this  study requires input images to have consistent sizes, so images need to be preselected. After standardized preprocessing, an additional method is required to handle image dimensions. Since structural and cellular metabolic changes in the tester's brain occur in random regions, restricting image sizes may lead to loss of critical information. To address this, multi-scale analysis methods such as pyramid pooling layers can be adopted.

(2) The global feature maps of brain segmentation images generated by the proposed network model can partially represent the influence of brain regions on diseases, serving as key areas for physicians to focus on during diagnosis. This can be developed as another research direction, which is not further explored in this  study.

\begin{IEEEbiography}[{\includegraphics[width=1in,height=1.25in,clip,keepaspectratio]{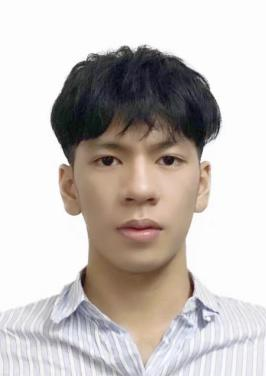}}]{Yang Ming} male, Han ethnicity, was born on October 24, 2003, in Guangzhou, Guangdong. In 2021, he enrolled in Guangdong Pharmaceutical University, majoring in Computer Science and Technology, and earned his Bachelor's degree in 2025.

During his university internship in 2025, he participated in research on visual algorithms for surgical robots at Zhuhai Hengqin Quanxing Medical Technology Co., Ltd. His specialized fields included 3D medical image segmentation of hepatic hemangiomas, multi-modal fusion and registration of liver ultrasound images with CT/MRI, and 3D brain MRI super-resolution reconstruction.

He has received several important awards: Outstanding Undergraduate Graduate of Guangdong Pharmaceutical University in 2025; Second Prize (Provincial First Prize) in the Artificial Intelligence Application category of the China University Student Computer Design Competition in 2024; and his project was approved in the National College Student Innovation and Entrepreneurship Training Program in 2024. His research interests focus on medical image algorithms and related computer science applications. 
\end{IEEEbiography}

\EOD

\end{document}